\documentclass[twocolumn,twocolappendix]{aastex631}
\usepackage{amsmath}

\shortauthors{Behmard et al.}
\graphicspath{{./}{figures/}}

\begin{document}

\title{Planet Engulfment Signatures in Twin Stars}

\author[0000-0003-0012-9093]{Aida Behmard}
\affiliation{Division of Geological and Planetary Sciences, California Institute of Technology, Pasadena, CA 91125, USA}

\author{Jason Sevilla}
\affiliation{Department of Astronomy, California Institute of Technology, Pasadena, CA 91125, USA}

\author[0000-0002-4544-0750]{Jim Fuller}
\affiliation{Department of Astronomy, California Institute of Technology, Pasadena, CA 91125, USA}


\begin{abstract}
Planet engulfment can be inferred from enhancement of refractory elements in the photosphere of the engulfing star following accretion of rocky planetary material. Such refractory enrichments are subject to stellar interior mixing processes, namely thermohaline mixing induced by an inverse mean-molecular-weight gradient between the convective envelope and radiative core. Using \texttt{MESA} stellar models, we quantified the strength and duration of engulfment signatures following planet engulfment. We found that thermohaline mixing dominates during the first $\sim$5$-$45 Myr post-engulfment, weakening signatures by a factor of $\sim$2 before giving way to depletion via gravitational settling on longer timescales. Solar metallicity stars in the 0.5--1.2 $M_{\odot}$ mass range have observable signature timescales of $\sim$1 Myr$-$8 Gyr, depending on the engulfing star mass and amount of material engulfed. Early type stars exhibit larger initial refractory enhancements but more rapid depletion. Solar-like stars ($M$ = 0.9$-$1.1 $M_{\odot}$) maintain observable signatures ($>$0.05 dex) over timescales of $\sim$20 Myr$-$1.7 Gyr for nominal 10 $M_{\oplus}$ engulfment events, with longer-lived signatures occurring for low-metallicity and/or hotter stars (1 $M_{\odot}$, $\sim$2$-$3 Gyr). Engulfment events occurring well after the zero-age main sequence produce larger signals due to suppression of thermohaline mixing by gravitational settling of helium (1 $M_{\odot}$, $\sim$1.5 Gyr). These results indicate that it may be difficult to observe engulfment signatures in solar-like stars that are several Gyr old.
\end{abstract}

\keywords{stars: abundances, planet-star interactions --- planetary systems, planets and satellites: dynamical evolution and stability --- planetary systems}


\section{Introduction} \label{sec:intro}
The formation and evolution of planetary systems can alter the primordial elemental abundances of planet host stars. For example, planet engulfment can produce refractory element enhancements within the engulfing star convective region due to ingestion of rocky planetary material (e.g., \citealt{oh2018}). However, such refractory enhancements may be weakened by internal mixing processes over time. While gravitational settling alone is not sufficiently rapid to diminish these enhancements (e.g., \citealt{pinsonneault2001}), thermohaline mixing may more effectively deplete overlying refractory material by allowing it to sink below the convective zone via ``metallic fingers" that stretch towards the deep stellar interior (e.g., \citealt{ulrich1972,kippenhahn1980,vauclair2004,theado2012,bauer2019ApJ}). As a form of double-diffusive mixing driven by heat diffusion, thermohaline mixing acts in the presence of a mean-molecular-weight ($\mu$) gradient, and is theorized to be particularly effective at removing accreted refractory material from the thin convective zones of hot stars due to more rapid cooling at the bottom of the convective envelope.

\citet{vauclair2004} was the first to consider thermohaline mixing in stellar interiors following accretion of planetary material. They analytically determined that thermohaline instabilities are capable of depleting accreted rocky matter out of convective envelopes on timescales of $\sim$1000 years, leaving only a small $\mu$-gradient at the end of the mixing process. Later work by \citet{garaud2011} presented numerical and semi-analytical estimations of thermohaline mixing using the coefficient derived by \citet{traxler2011} from empirical fitting of 3D numerical simulation results, and found that the refractory depletion timescale following engulfment of a 1 $M_{\textrm{J}}$ planet depends strongly on increasing stellar mass, with enhancements dropping to 10\% within 600 Myr for a 1.3 $M_{\odot}$ star, 60 Myr for a 1.4 $M_{\odot}$ star, etc.

More recently, \citet{theado2012} modeled engulfment with planet masses ranging from 1 $M_{\oplus}$ to 1.5 $M_{\textrm{J}}$ using the Toulouse-Geneva Evolution Code \citep{richard2004,huibonhoa2008} considering thermohaline mixing and atomic diffusion. They found that the $\mu$-gradient is softened on timescales of a few to tens of Myr depending on the accretion scenario, and noted that their depletion timescale is much larger than those of previous studies because they correctly assumed the characteristic mixing length to be the entire mixed region rather than just the ``metallic finger" dimensions. 
\citet{theado2012} showed how different mixing prescriptions could lead to different results, especially in the case of a small composition gradient (near the minimum for thermohaline instability to operate), as often occurs in models of planet engulfment.
This highlights the need for stellar models that include all relevant interior processes for accurately constraining depletion timescales. Such timescales are important for investigations of engulfment because they determine how long engulfment signatures remain observable, which is necessary for placing engulfment events within planetary system histories. 


Stellar abundance patterns suggestive of planet engulfment have been observed in several systems, indicating that engulfment may be more common than predictions from theory suggest (Galactic occurrence rates of  0.1--1 yr$^{-1}$, \citealt{metzger2012}). More specifically, there are several individual binary systems reported in the literature with significant ($>$0.05 dex\footnote{In this work, we adopt the standard ``bracket" chemical abundance notation [X/H] = $A$(X) - $A$(X)$_{\odot}$, where $A$(X) = log($n_{\textrm{X}}$/$n_{H}$) + 12 and $n_{\textrm{X}}$ is the number density of species X in the star's photosphere.}) refractory differences between the two stars  \citep{ramirez2011,mack2014,tucci_maia2014,teske2015,ramirez2015,biazzo2015,saffe2016,teske2016,adibekyan2016,saffe2017,tucci_maia2019,ramirez2019,nagar2020,galarza2021,jofre2021}. Such abundances differences between stellar siblings suggest post-birth chemodynamical processes such as engulfment because bound stellar companions are born from the same natal gas cloud and thus assumed to be chemically homogeneous at birth to within $\sim$0.05 dex (e.g., \citealt{de_silva2009}). Larger binary samples exhibit abundance differences as well. For example, \citet{hawkins2020} assessed 25 comoving, wide binaries and found that 5 systems exhibit $\Delta$[Fe/H] $\sim$ 0.10 dex. 

To determine how planet engulfment signatures are affected by stellar mixing processes, namely thermohaline instabilities, we ran tests with the \texttt{MESA} (Modules for Experiments in Stellar Astrophysics) stellar evolution code \citep{paxton2011,paxton2013,paxton2015,paxton2018,paxton2019}. Our \texttt{MESA} stellar models and implementation of non-standard mixing processes such as thermohaline instabilities
are outlined in Section \ref{sec:model}. The results of our \texttt{MESA} models considering different engulfment conditions are presented in Section \ref{sec:results}. We discuss the implications of these results for planet engulfment signatures in Section \ref{sec:discussion}, and provide a summary in Section \ref{sec:summary}.

\section{Stellar Models} \label{sec:model}
We computed our stellar models using the open-source 1D stellar evolution code \texttt{MESA}. These models are non-rotating with masses of 0.5--1.2 $M_{\odot}$ and solar metallicities of $Z$ = 0.017. We did not include more massive stars, where gravitational settling and radiative levitation can potentially produce huge changes in surface composition, but are complicated by poorly understood additional mixing processes not included in our models (e.g., Eddington-Sweet circulation and wave mixing).

Our modeling procedure follows that of \citet{sevilla2022}. In brief, the models were run in three stages, where the first stage evolved the stars up to the zero-age main sequence (ZAMS), the second stage simulated planet engulfment of a 1, 10, or 50 $M_{\oplus}$ planet via accretion of bulk-Earth composition material, and the third stage evolved the stars up to the end of their main sequence (MS) lifetimes. We then computed the refractory species mass fraction remaining in their convective zones as a function of stellar age to assess the timescales over which engulfment signatures remained observable. Throughout these \texttt{MESA} runs we applied relevant mixing processes, namely convective overshoot, elemental diffusion, radiative levitation, thermohaline instabilities, and additional mixing required to reproduce observations of surface lithium abundances. These processes are discussed in more detail below. 

\subsection{Input Physics}
We applied convective overshoot via the prescription taken from \citet{herwig2000}, with \texttt{MESA} implementation detailed in \citet{paxton2011}:
\begin{eqnarray}
D_{\rm ov} = D_{\rm conv, 0}\hspace{0.7mm} \exp  \left ( -\frac{2z}{f \lambda_{P, 0}}\right) , \hspace{1mm}
\end{eqnarray}
where $D_{\rm conv, 0}$ is the diffusion coefficient at a location $f_0$ scale heights inside the convective zone, $\lambda_{P, 0}$ is the pressure scale height at that location, $z$ is the distance away from this location, and $f$ is an adjustable parameter that sets the characteristic size of the region undergoing convective overshoot.
We set $f$ and $f_0$ to 0.02 and 0.005, respectively.

We implemented elemental diffusion via the prescription detailed in \citet{paxton2015} and \citet{paxton2018} that solves the Burgers' equations in a similar fashion to \citet{thoul1994}. In \texttt{MESA}, diffusion is carried out for a set of chemical species organized in $``$classes$"$, where a single class may contain several isotopes, and a chosen representative isotope is used to solve the diffusion equations and set the diffusion velocities for all species in that class. In our case, we created individual classes for each of the 13 most abundant species included in our bulk-Earth accretion that simulates engulfment, so each species is its own representative isotope. 

We also applied radiative levitation according to \citet{hu2011}, implemented in \texttt{MESA} as described in \citet{paxton2015}. The \citet{hu2011} prescription is based on that of \citet{thoul1994}, with a few modifications. These include an additional force term in the basic diffusion equations that describes the radiative acceleration on a chemical species as a function of its average ion charge.
We only applied radiative levitation to our stellar models with masses of 0.7 $M_{\odot}$ and above. This is justified because radiative levitation dominates atomic diffusion for stars with temperatures above 6000 K (e.g., \citealt{richer1998, deal2020, campilho2022}), and is not expected to operate efficiently in low mass stars.

Thermohaline mixing acts in the presence of a mean molecular weight ($\mu$) inversion in regions that satisfy the Ledoux criterion for convective stability, such that
\begin{eqnarray}
\nabla_{T} - \nabla_{\textrm{ad}} \leq B \leq 0 .
\end{eqnarray}
Here, $B$ is the Ledoux term that accounts for composition gradients (e.g., \citealt{brassard1991}), and $\nabla_{T}-\nabla_{\textrm{ad}}$ is the difference between the actual temperature gradient and the adiabatic temperature gradient, also referred to as the superadiabaticity. We included thermohaline mixing in our \texttt{MESA} models according to the prescription of \citet{brown2013} derived via 3D numerical simulations and linear stability analyses. This prescription is more accurate than previous thermohaline implementations, such as that of \citet{kippenhahn1980} which overestimates mixing efficiency in certain fingering convection configurations \citep{prat2015}. It also corrects inconsistencies in the models of \citet{denissenkov2010} and \citet{traxler2011}.




There are other mixing processes at play that we did not explicitly model because their effects are poorly understood, e.g., rotationally induced mixing. To account for this, we included a \texttt{min\_D\_mix = 700} command in our \texttt{MESA} inlist. This amount of extra mixing allowed us to reproduce the observed lithium abundances for $\sim$1 $M_{\odot}$ stars from \citet{sestito2005}. For more details on the comparison with lithium observations, see \citet{sevilla2022}.

\subsection{Bulk-Earth Accretion} \label{sec:bulk_earth_accretion}

\begin{figure*}[t]
\centering
    \includegraphics[width=1\textwidth]{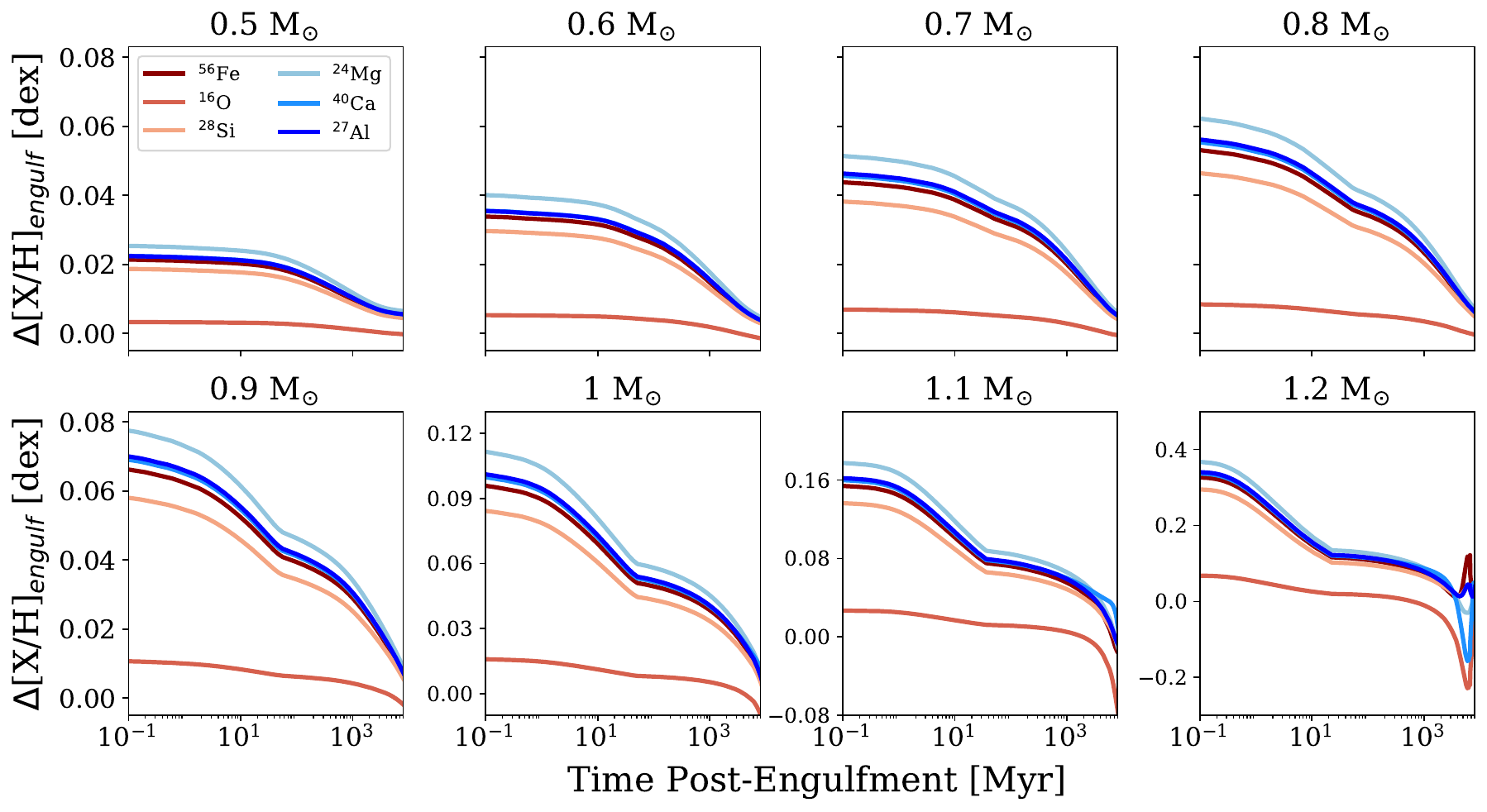}
\caption{Post-engulfment abundances for the top six most abundant bulk-Earth elements over time for our complete stellar mass model range (0.5$-$1.2 $M_{\odot}$) following near-instantaneous accretion of a 10 $M_{\oplus}$ bulk-Earth composition planet. The abundances of comparison \texttt{MESA} models with no accretion were subtracted off. Each element is represented by its most abundant isotope because \texttt{MESA} chemical networks are written in terms of isotopic species.}
\label{fig:figure1}
\end{figure*}

To simulate planet engulfment, we began accreting bulk-Earth composition material once the \texttt{MESA} stellar models reached ZAMS, consisting of the top 13 most abundant elements \citep{mcdonough2003}. 
For a complete analysis of the lithium enrichment evolution, see \citet{sevilla2022}. Because \texttt{MESA} chemical networks are written in terms of isotopic species, we accreted the most abundant isotope of each element included in our bulk-Earth accretion scheme.

\texttt{MESA} does not allow for instantaneous engulfment of a planetary mass body. Instead, planet engulfment was simulated through rapid accretion of bulk-Earth composition material. We chose accretion rates that ensured the total planetary mass was accreted within 10,000 years for all combinations of 1, 10, and 50 $M_{\oplus}$ engulfed masses and host stars of 0.5$-$1.2 $M_{\odot}$. An accretion period of 10,000 years is appropriately short for simulating engulfment events, and should not affect our results because our observed refractory depletion occurs on timescales of $>$1 Myr (Figure \ref{fig:figure1}). 

We also tested two alternative accretion prescriptions that mimic late heavy bombardment (LHB) and disk accretion scenarios. LHB-like accretion begins when the star reaches an age of 500 Myr and lasts for 300 Myr (e.g., \citealt{bottke2017}), while disk accretion begins 10 Myr after the star is born and lasts for 3 Myr, the nominal disk lifetime for MS stars (e.g., \citealt{hartmann2016}). We tested accretion of 1 and 10 $M_{\oplus}$ of planetary material for both scenarios. 
After engulfment according to near-instantaneous, LHB, or disk accretion scenarios, we evolved the stellar models to the ends of their MS lifetimes. We utilized the default definition for the end of the MS, which is when the core hydrogen mass fraction drops below 10$^{-10}$.

\begin{figure*}[t]
\centering
    \includegraphics[width=1\textwidth]{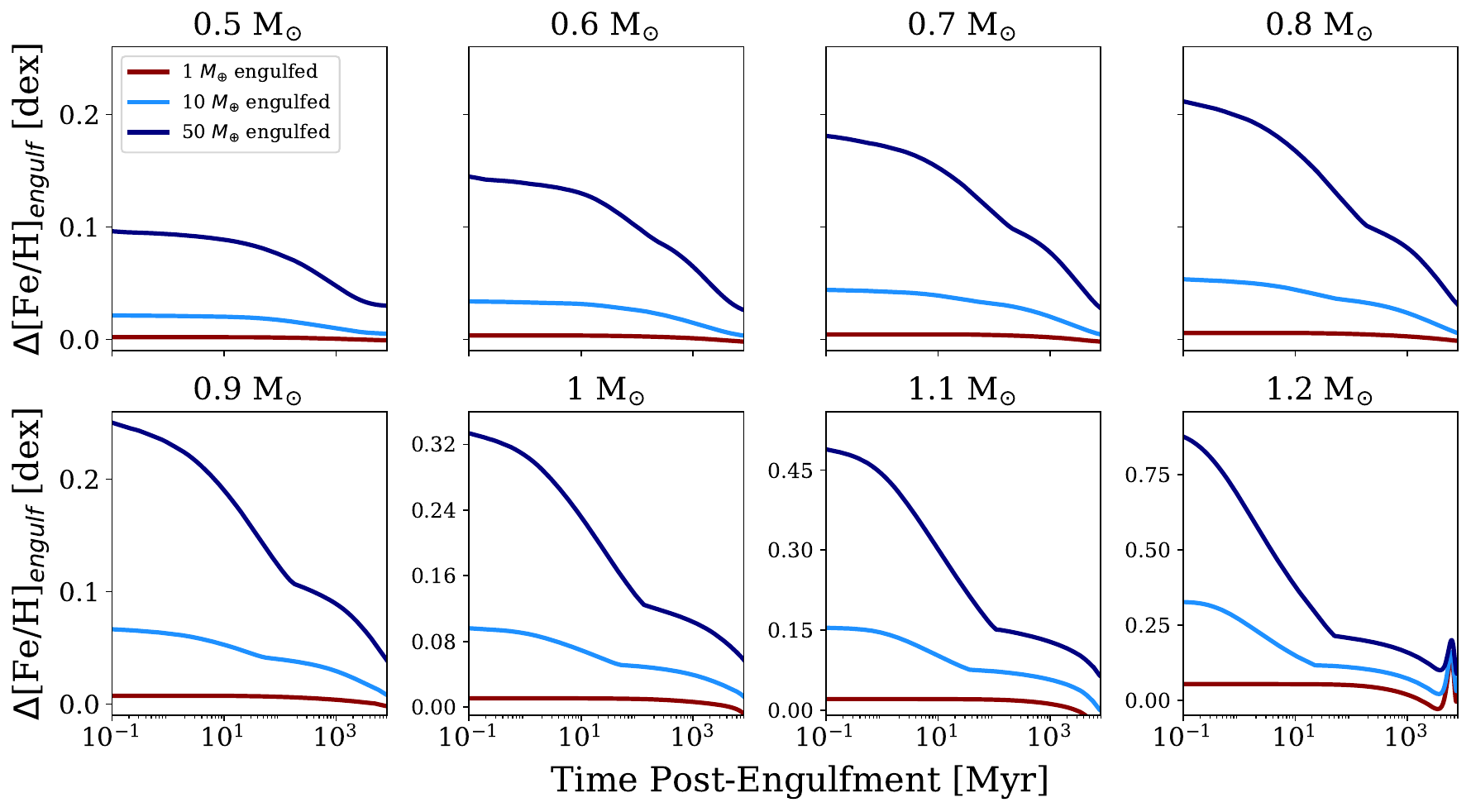}
\caption{Post-engulfment [Fe/H] abundances over time for our complete stellar mass model range (0.5$-$1.2 $M_{\odot}$) following near-instantaneous accretion of a 1, 10, or 50 $M_{\oplus}$ bulk-Earth composition planet, marked in red, blue, and navy, respectively. The abundances of comparison \texttt{MESA} models with no accretion were subtracted off.}
\label{fig:figure2}
\end{figure*}


\section{Results}
\label{sec:results}

\subsection{Near-Instantaneous Engulfment}
\label{sec:near_instaneous_engulfment}
We first examined how refractory enrichment resulting from engulfment changes as a function of engulfed planet mass and stellar type. This was carried out by running \texttt{MESA} models that included accretion of 1, 10, or 50 $M_{\oplus}$ of planetary material with bulk-Earth compositions, and models with no accretion as comparison points. We took the differential abundances between the engulfment and non-engulfment models as our potential engulfment signatures. These differential measurements mimic the signatures found in observations because refractory enrichments from engulfment are detected with respect to pristine, chemically homogeneous stars, e.g., a bound companion or fellow cluster member. 

As shown in Figure \ref{fig:figure1}, for low-mass stars (0.5$-$0.7 $M_{\odot}$), engulfment of a 10 $M_{\oplus}$ planet does not produce observable enrichment at $>$0.05 dex levels because the accreted material is heavily diluted within the deep stellar convective envelope. The enrichment is more pronounced for more massive stars with thinner convective envelopes; for solar-like stars (0.8$-$1.2 $M_{\odot}$), engulfing 10 $M_{\oplus}$ of planetary material initially produces enrichment at levels of $\sim$0.06$-$0.33 dex. Because the 0.8$-$0.9 $M_{\odot}$ stars still have comparatively deep envelopes, the initial enrichment is not well above 0.05 dex to begin with, and drops below this after by $\sim$20 Myr. However the higher mass stars in this range (1$-$1.2 $M_{\odot}$) maintain enrichment at $>$0.05 dex for $\sim$90 Myr$-$2 Gyr.

Figure \ref{fig:figure1} shows that for models with higher mass ($M \gtrsim 1 \, M_\odot$), the refractory abundances of a star that engulfed a planet are predicted to eventually decrease \textit{below} those of a star that did not engulf a planet within 8 Gyr post-engulfment. The strength of this counter-intuitive effect increases with stellar mass. Inspection of our $1.2 \, M_\odot$ models with engulfment reveals that they are typically $\sim$200 K cooler and $\sim$1.5\% larger in radius than their reference models. Surprisingly, the models with engulfment have thinner surface convection zones, leading to faster gravitational settling of heavy elements and hence lower refractory abundances long after the engulfment event. The exact cause of this structural difference is not clear and should be examined in future work. There is an exception for certain isotopic species that are subject to radiative levitation (e.g, $^{56}$Fe, $^{27}$Al, $^{28}$Si), whose abundances increase at $\sim$5 Gyr in the $1.2 \, M_\odot$ model (Figures \ref{fig:figure1} and \ref{fig:figure2}).


Apart from this case, we found little difference between the chemical species in our models; the relative abundance of each decreases on nearly identical timescales. For thermohaline mixing, this is expected as it affects all elements equally. In principle, gravitational settling causes heavier elements to sink below the convective zone faster, but the difference is not discernible in our results. Note that the enrichment signature provided by oxygen is much weaker than those of other elements due to the large amount of oxygen already present in the star.

\subsection{Dependence on Planet Mass}
We also examined near-instantaneous engulfment events of 1 and 50 $M_{\oplus}$ of planetary material (Figure \ref{fig:figure2}). For 1 $M_{\oplus}$ engulfment, refractory depletion behavior across different stellar mass regimes is similar to that of 10 $M_{\oplus}$ engulfment, but the initial enrichment levels are lower. Consequently, stars with masses in the range 0.5$-$1.1 $M_{\odot}$ do not exhibit enrichment above $\sim$0.05 dex immediately post-engulfment, rendering their engulfment signatures undetectable. The 1.2 $M_{\odot}$ model begins with $\sim$0.05 dex enrichment and sustains this level of enrichment for $\sim$100 Myr. 



For 50 $M_{\oplus}$ engulfment, the initial enrichment is comparatively high as expected from accretion of more planetary material. Stars in the 0.5$-$0.7 $M_{\odot}$ mass regime begin with enrichment at 0.10$-$0.19 dex levels, and maintain enrichment above $\sim$0.05 dex for $\sim$900 Myr$-$3 Gyr. Solar-like stars (0.8$-$1.2 $M_{\odot}$) begin with enrichment at $\sim$0.25$-$1.0 dex, and maintain $\sim$0.05 dex enrichment on timescales of $\sim$4$-$8 Gyr. 


\subsection{LHB and Disk Accretion}
\label{sec:LHB_disk}
\begin{figure}[t]
\centering
    \includegraphics[width=0.49\textwidth]{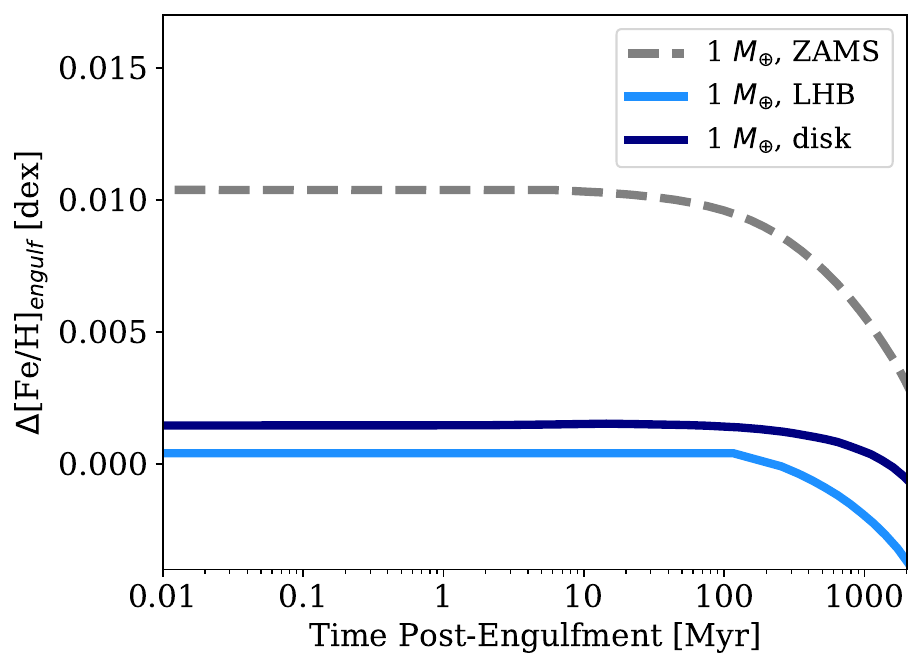}
\caption{Post-engulfment [Fe/H] abundances over time for a 1 $M_{\odot}$ \texttt{MESA} model that underwent accretion of a 1 $M_{\oplus}$ planet under different conditions: the standard post-ZAMS near-instantaneous engulfment scenario (gray, dashed), an LHB-like scenario (blue), and a disk accretion-like scenario (navy). The abundances of comparison \texttt{MESA} models with no accretion were subtracted off.}
\label{fig:figure3}
\end{figure}

The LHB is considered to be the last dramatic dynamical event in the history of the Solar System, when giant planet migration disrupted the orbits of inward-lying planetesimals and perhaps shepherded some into the Sun. We modeled LHB-like events to compare how the resulting engulfment signatures may differ from those of engulfment at ZAMS. 
To mimic LHB-like accretion, we ran a 1 $M_{\odot}$ model undergoing gradual accretion of 1 $M_{\oplus}$ of bulk-Earth composition planetesimals from a stellar age of 500$-$800 Myr, thus beginning $\sim$470 Myr after ZAMS. We found that engulfment signatures from an LHB-like event are almost completely depleted post-engulfment because refractory material is continuously depleted over the longer accretion period. The engulfment signature after LHB-like accretion is complete is shown in Figure \ref{fig:figure3}.

\begin{figure*}[t]
\centering
\begin{minipage}{0.47\textwidth}
  \centering
  \includegraphics[width=0.99\linewidth]{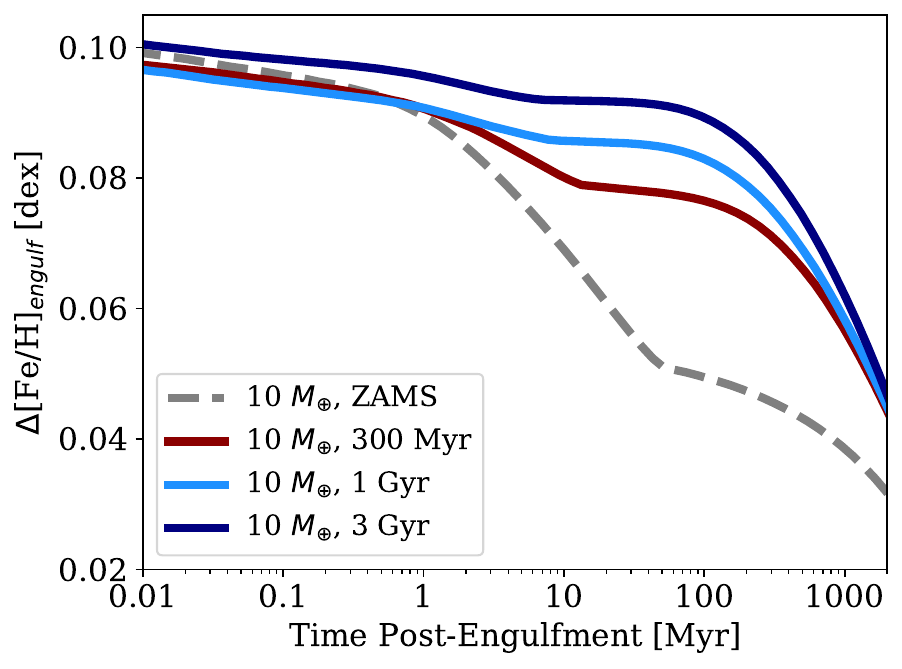}
  \caption{Post-engulfment [Fe/H] abundances  over time for a 1 $M_{\odot}$ \texttt{MESA} model that underwent accretion of a 10 $M_{\oplus}$ planet at different times: ZAMS (gray, dashed), 300 Myr post-ZAMS (red), 1 Gyr post-ZAMS (blue), and 3 Gyr post-ZAMS (navy).
The abundances of comparison \texttt{MESA} models with no accretion were subtracted off.}
  \label{fig:figure4}
\end{minipage}%
\hspace{9mm}
\begin{minipage}{0.465\textwidth}
  \centering
  \vspace{-4.4mm}
  \includegraphics[width=0.99\linewidth]{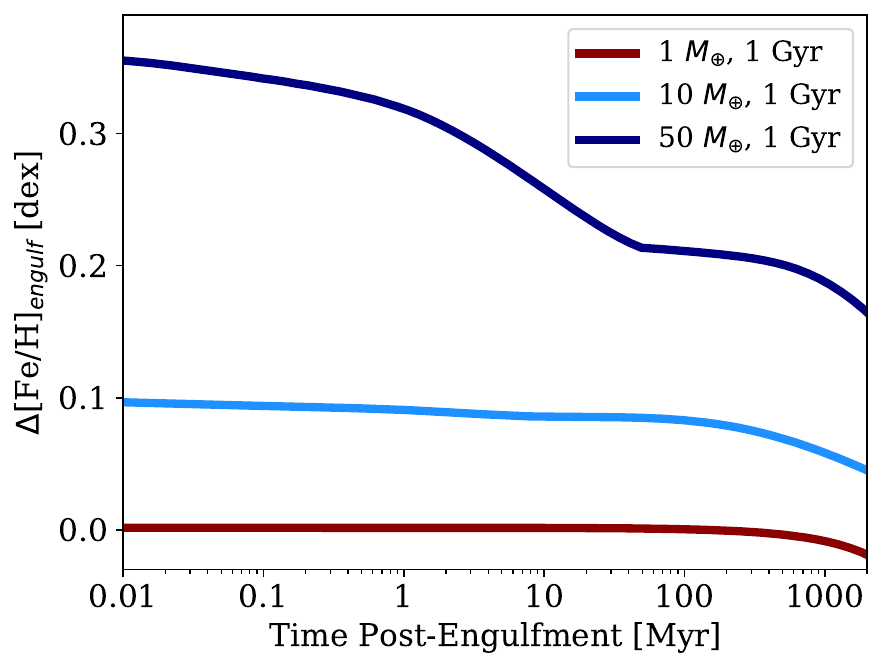}
  \caption{Post-engulfment [Fe/H] abundances over time for a 1 $M_{\odot}$ \texttt{MESA} model that underwent accretion of a 1 (red), 10 (blue), or 50 (navy) $M_{\oplus}$ planet at 1 Gyr post-ZAMS. The abundances of comparison \texttt{MESA} models with no accretion were subtracted off.}
  \label{fig:figure5}
\end{minipage}
\end{figure*}

We were also interested in modeling a protoplanetary disk accretion scenario, which by definition occurs while the disk is still present. To mimic these conditions, we began accretion once the star reached an age of 10 Myr, for a duration of 3 Myr as mentioned in Section \ref{sec:bulk_earth_accretion}. At 10 Myr, the star is  still in the pre-MS phase, $\sim$20 Myr away from reaching ZAMS and almost fully convective. Thus, any accreted material will become heavily diluted within a large volume of the stellar interior. We tested accretion of 1 and 10 $M_{\oplus}$, and found that the resulting engulfment signatures are slightly stronger than those resulting from LHB scenarios, but weaker compared to those of near-instantaneous engulfment scenarios because the refractory material is comparatively diluted (Figure \ref{fig:figure3}).


\subsection{Late-Stage Engulfment}

\begin{figure*}[t]
    \centering
    \begin{minipage}{0.495\textwidth}
        \centering
        \includegraphics[width=0.99\textwidth]{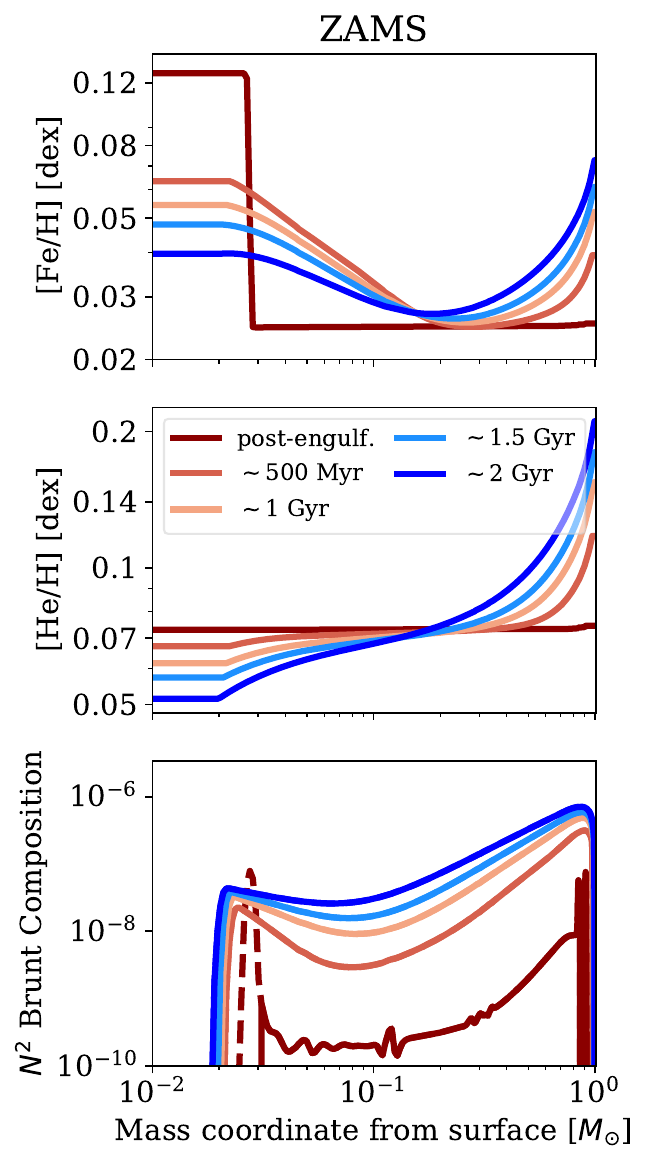} 
    \end{minipage}\hfill
    \begin{minipage}{0.495\textwidth}
        \centering
        \includegraphics[width=0.99\textwidth]{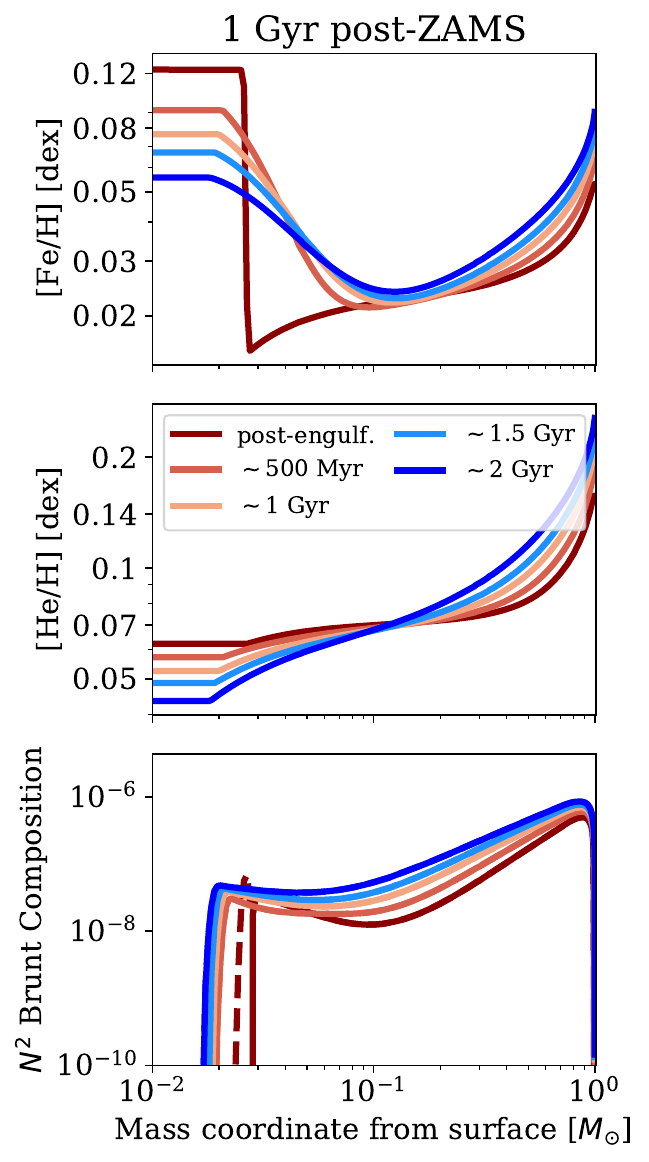} 
    \end{minipage}
    \caption{Plots of [Fe/H], [He/H], and $N^2_{\rm comp}$ as a function of mass coordinate for a $1 \, M_\odot$ \texttt{MESA} stellar model. Negative $N^2_{\rm comp}$ values are shown via dashed line (absolute values are used to enable visualization in log scale). The left panel corresponds to the case of a 1 $M_{\odot}$ model engulfing 10 $M_{\oplus}$ of planetary material exactly at ZAMS, while the right panel corresponds to an equivalent case, but with engulfment occurring at 1 Gyr post-ZAMS. We sampled these quantities at five different time points (measured relative to the ZAMS age) from engulfment to an age of $\sim$2 Gyr, as depicted by different line colors.}
\label{fig:figure6}
\end{figure*}

Finally, we tested near-instantaneous engulfment occurring at 300 Myr, 1 Gyr, or 3 Gyr post-ZAMS rather than exactly at ZAMS to examine refractory enrichment evolution from late-stage accretion. For these cases we considered engulfment of a 10 $M_{\oplus}$ mass by a 1 $M_{\odot}$ star. Notably, later engulfment results in systematically larger surface refractory enhancements; for the 3 Gyr model, the enhancement is $\sim$1.5$-$2 times that of the ZAMS accretion model at $\gtrsim$10 Myr post-engulfment times (Figure \ref{fig:figure4}). 
This effect is also shown in the trends of different engulfed masses; we tested engulfment of a 1, 10, or 50 $M_{\oplus}$ bulk-Earth planetary companion by a 1 $M_{\odot}$ star at 1 Gyr post-ZAMS (Figure \ref{fig:figure5}), and found that the 1 Gyr post-ZAMS cases all exhibit slower depletion of surface refractory enhancement compared to ZAMS engulfment (Figure \ref{fig:figure2}).

The primary reason for the larger signal in late-stage engulfment is a smaller amount of thermohaline mixing, which results in less depletion after engulfment as shown in Figures \ref{fig:figure4} and \ref{fig:figure5}. The weakened thermohaline mixing arises due to its interaction with gravitational settling. As stars age, gravitational settling slowly creates a stabilizing $\mu$-gradient as helium settles downwards. This inhibits thermohaline mixing when planetary material is engulfed, thus weakening refractory depletion and leading to longer-lived engulfment signatures. This effect is discussed further in \citet{theado2012} and \citet{sevilla2022}.

To further illustrate how this manifests in our \texttt{MESA} models, we plotted [Fe/H], [He/H], and the compositional component of the Brunt-V\"ais\"al\"a frequency ($N^2_{\rm comp}$) as a function of mass coordinate for the 1 $M_{\odot}$ model with 10 $M_{\oplus}$ engulfment considering both ZAMS and 1 Gyr post-ZAMS engulfment (Figure \ref{fig:figure6}).
The [Fe/H] panels of Figure \ref{fig:figure6} showcase that some gravitational settling has already occurred in the 1 Gyr post-ZAMS model compared to the ZAMS model, increasing the value of [Fe/H] at larger depths.
Similarly, helium settling creates a stabilizing composition gradient beneath the convection zone, as can be seen by the increasing value of [He/H] and [Fe/H] with depth below the convective zone in the post-engulfment models (though note that nuclear burning also increases [He/H] near the core).

We further examined this effect on the $\mu$-gradient by plotting $N^2_{\rm comp}$, a measure of stability to convection in a fluid medium. A positive $N^2_{\rm comp}$ indicates a stable $\mu$-gradient, while a negative value
indicates that thermohaline mixing can occur. In both models, the bottom panels of Figure \ref{fig:figure6} show that $N^2_{\rm comp}$ is negative just below the convective zone of the post-engulfment model. However, at larger depths, $N^2_{\rm comp}$ has lower values for engulfment at ZAMS compared to engulfment at 1 Gyr post-ZAMS, which indicates that the 1 Gyr post-ZAMS $\mu$-gradient is more stable to begin with. Thermohaline mixing cannot penetrate as deeply into the interior in the 1 Gyr post-ZAMS model, which thus experiences less refractory depletion via thermohaline mixing over time. 

\begin{figure}[t]
\centering
    \includegraphics[width=0.48\textwidth]{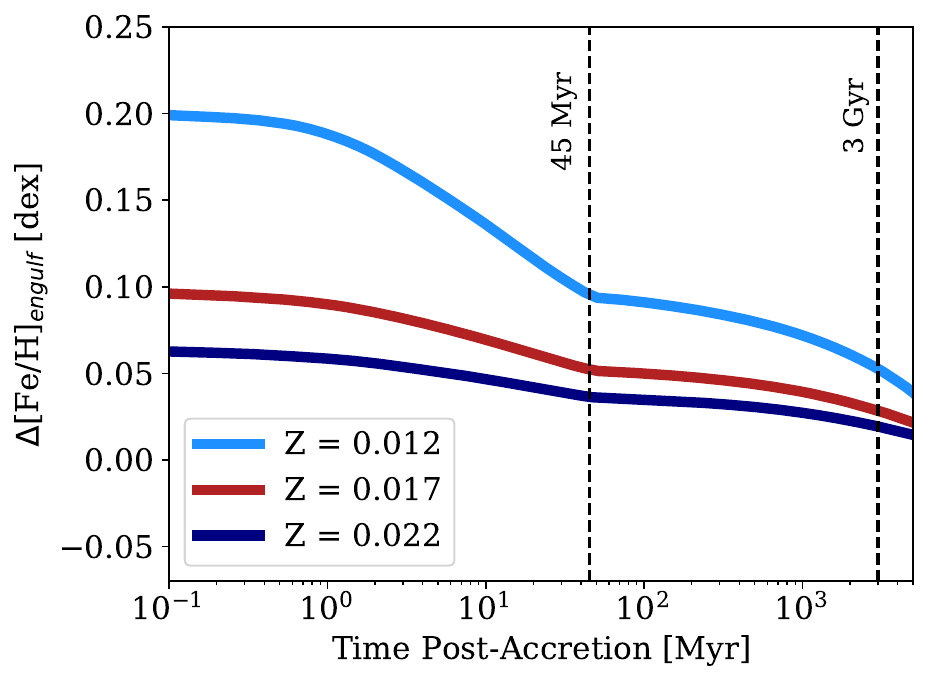}
\caption{Post-engulfment [Fe/H] abundances
over time for 1 $M_{\odot}$ models following near-instantaneous accretion of a 10 $M_{\oplus}$ bulk-Earth composition planetary companion at ZAMS. The three 1 $M_{\odot}$ models have different metallicities of slightly sub-solar ($Z$ = 0.012, blue), solar ($Z$ = 0.017, red), and slightly super-solar ($Z$ = 0.022, navy). The abundances of comparison \texttt{MESA} models with no accretion were subtracted off. The refractory enhancements drop by a factor of $\sim$2 after $\sim$45 Myr. For the sub-solar metallicity model, observable signatures last for $\sim$3 Gyr.}
\label{fig:figure7}
\end{figure}

\subsection{Engulfing Star Metallicity}
\label{sec:metallicity}

While we defaulted to solar metallicities ($Z$ = 0.017) in our \texttt{MESA} models, we also examined how engulfing star metallicities affect engulfment signatures. We ran 1 $M_{\odot}$ models of $Z$ = 0.012 and 0.022 with accretion of 10 $M_{\oplus}$ planets, and found that engulfment signature strength is strongly inversely correlated with host star metallicity (Figure \ref{fig:figure7}). This is due to two effects: lower metallicity stars have fewer metals leading to stronger relative refractory enrichments, and lower metallicity stars have thinner convective envelopes. The contributions of both these factors to convective zone metal content before engulfment is
\begin{eqnarray}
M_{Z,\rm CZ} = f_{Z,\rm CZ}\hspace{0.7mm} M_{\rm CZ} , \hspace{1mm}
\end{eqnarray}
where $f_{Z,\rm CZ}$ is the mass fraction of refractory species in the convective zone, and $M_{\rm CZ}$ is the convective zone mass. We computed $M_{\rm CZ}$ for the three metallicity cases to be $\sim$0.014 $M_{\odot}$, $\sim$0.024 $M_{\odot}$, and $\sim$0.030 $M_{\odot}$ for $Z$ = 0.012, 0.017, and 0.022, respectively. The convective zone refractory mass fraction $f_{Z,\rm CZ}$ can be approximated as $Z$, so $M_{\rm CZ}$ increases slightly faster with metallicity than $f_{Z,\rm CZ}$, but both factors are nearly equally important for setting the pre-engulfment metal content. Because the total metal content in the convective envelope is much larger at higher metallicity, an engulfed planet has a proportionally smaller effect. 


Notably, the lowest metallicity model with $Z$ = 0.012 exhibits an engulfment signature that remains above the 0.05 dex observable level for $\sim$3 Gyr. This implies that stars with sub-solar metallicities may exhibit engulfment signatures that are fairly long lived. For context, the $Z$ distribution of $\sim$300 G dwarfs in the Solar neighborhood ranges from $\sim$0.001$-$0.043, and peaks at $\sim$0.009 dex \citep{rocha1996}. A more recent study of $\sim$17,000 K and $\sim$24,000 G dwarfs outside the Solar vicinity at 0.2$-$2.3 kpc from the Galactic plane derived similar $Z$ distributions, ranging from $\sim$0.0002$-$0.03 and peaking at $\sim$0.01 \citep{schlesinger2012}. Thus, the distribution of solar-like stars in our neighborhood leans relatively metal-poor, indicating that observable engulfment signatures may persist on $\sim$3 Gyr timescales for many nearby $\sim \! 1 \, M_\odot$ stars.

\begin{figure*}[t]
\centering
    \includegraphics[width=0.99\textwidth]{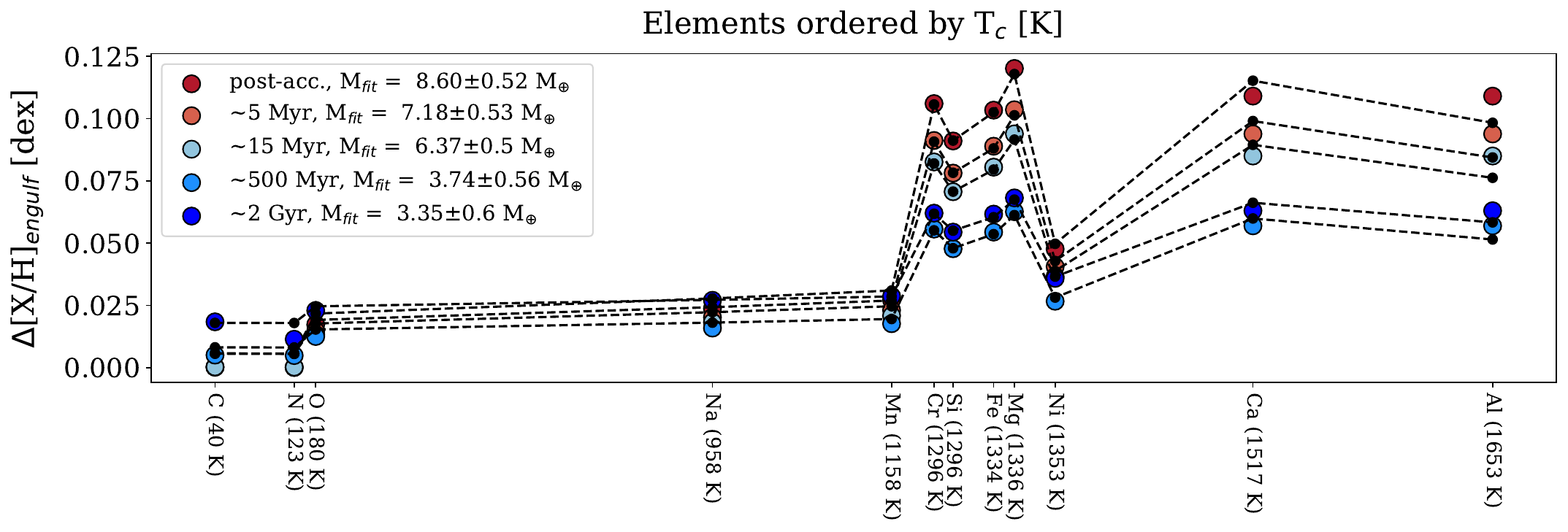}
    \vspace{-2mm}
\caption{Differential abundances between a 1 $M_{\odot}$ model and a 1 $M_{\odot}$ model that accreted 10 $M_{\oplus}$ of bulk-Earth composition material (colored circles) for several post-engulfment times. The abundances are ranked by $T_{c}$ of elements for solar-composition gas from \citet{lodders2003}.
The modeled abundances are represented by the black dots. The amount of apparent engulfed material at each time is provided in the upper left corner of the plot.}
\label{fig:figure8}
\end{figure*}

\subsection{Binary Observation Simulations}
\label{sec:binary_observations}

As mentioned in Section \ref{sec:near_instaneous_engulfment}, engulfment signatures are detected in observations as patterns in the differential abundances of an engulfing star and a comparison stellar companion. The characteristic pattern indicative of an engulfment event is a trend of increasing differential abundances as a function of the elemental condensation temperature $T_{c}$ (e.g., \citealt{oh2018}). We expect a $T_{c}$-dependent abundance pattern following engulfment; elements with higher $T_{c}$ are more likely to remain in the condensed phase throughout the disk and become locked in solid planetary material. Thus, planet compositions will exhibit higher abundances of species in order of $T_{c}$, which will be reflected in the photosphere of a star that has engulfed a planet.  

We simulated how engulfment signatures would appear in binary system observations by computing differential abundances for elements that span a range of $T_{c}$ between two G-type (1 $M_{\odot}$) stars, one of which underwent engulfment of a 10 $M_{\oplus}$ bulk-Earth composition planet at ZAMS. The differential abundances from \texttt{MESA} display the characteristic $T_{c}$ trend that signifies engulfment (Figure \ref{fig:figure8}, colored points). We also constructed an engulfment model similar to that of \citet{oh2018} for estimating the mass of bulk-Earth composition \citep{mcdonough2003} material engulfed in one star given abundance measurements for a binary pair. This model estimates the apparent engulfed mass, which is the quantity inferred from stellar surface measurements assuming the planetary material is mixed throughout the convective zone, but not deeper.

Assuming solar abundances for the engulfing star [X/H] \citep{asplund2009}, the model computes the mass fraction of each element X as follows:
\begin{eqnarray}
f_{\textrm{X,photo}} = \frac{10^{[\textrm{X/H}]} m_{\textrm{X}}}{\Sigma_{\textrm{X}} 10^{[\textrm{X/H}]} m_{\textrm{X}}} \hspace{0.5mm},
\end{eqnarray}
where $m_{\textrm{X}}$ is the mass of each element. Given a total mass of accreted material $M_{\textrm{acc}}$ and accreted mass fractions for each element $f_{\textrm{X,acc}}$, the abundance difference between the stars is

\begin{eqnarray}
\Delta[\textrm{X/H}] = \textrm{log}_{10} \frac{f_{\textrm{X,photo}} \hspace{0.4mm} f_{\textrm{CZ}} \hspace{0.4mm} M_{\textrm{star}} \hspace{0.4mm}+ \hspace{0.4mm} f_{\textrm{X,acc}} \hspace{0.4mm} M_{\textrm{acc}}}{f_{\textrm{X,photo}} \hspace{0.4mm} f_{\textrm{CZ}}\hspace{0.4mm} M_{\textrm{star}}} \hspace{0.5mm},
\end{eqnarray}

\noindent where $f_{\textrm{CZ}}$ is the mass fraction of the outer convective zone. For more details on the engulfment model, see \citet{oh2018}. Our implementation makes use of the \texttt{dynesty} nested sampling code \citep{speagle2020} to fit the estimated amount of engulfed mass. We also added abundance scatter estimated as a function of observation signal-to-noise ratio (SNR) as reported by \citet{brewer2018}, assuming an SNR level of 200/pix to mimic high quality observations.

We fit for the apparent amount of engulfed mass according to our model at five different time points that span from immediately post-engulfment to $\sim$2 Gyr after the engulfment event (Figure \ref{fig:figure8}, black points connected via dashed line). The fitted mass begins at 8.60 $\pm$ 0.52 $M_{\oplus}$ and falls to 3.35 $\pm$ 0.60 $M_{\oplus}$ over this time period. Our model underestimates the initial engulfed mass because it assumes all accreted material is initially contained in the convective zone, which is not completely accurate; overshoot mixing pulls engulfed material underneath the convective zone to an additional depth of $\sim$10\% the convective zone mass. We note that while the signature appears clear even at the $\sim$2 Gyr post-engulfment point, abundance measurements from real observations will exhibit scatter due to instrumental effects, observing conditions, etc., making refractory enhancements below the 0.05 dex level at this point nearly undetectable.

\section{Discussion} \label{sec:discussion}
The strength and duration of planet engulfment signatures are affected by the amount of mass engulfed, the properties of the engulfing star, and the conditions of the engulfment event, e.g., when the event occurred within the lifetime of the system. We discuss how these factors influence engulfment signature evolution here.

We considered engulfed masses of 1, 10, and 50 $M_{\oplus}$. 10 $M_{\oplus}$ engulfment can be considered nominal as formation of a solid $\sim$10 $M_{\oplus}$ core triggers runaway gas accretion according to the core accretion model of planet formation. However some giant planets may possess up to $\sim$100 $M_{\oplus}$ worth of refractory material, perhaps due to protoplanet merger events that occurred prior to disk outgassing \citep{ginzburg2020}. We tested planet masses up to 50 $M_{\oplus}$ to account for occasional engulfment of such $``$heavy metal Jupiters$"$. Larger amounts of engulfed mass result in higher initial enrichment, but also stronger $\mu$-gradients that enhance thermohaline mixing and subsequent refractory depletion. These effects are illustrated throughout different engulfing star mass regimes (Figure \ref{fig:figure2}). For low-mass stars (0.5$-$0.7 $M_{\odot}$), observable enrichment is only achieved for engulfed planetary masses of $\geq$50 $M_{\oplus}$, and last for $\sim$900 Myr$-$3 Gyr. For more massive, solar-like (0.8$-$1.2 $M_{\odot}$) stars with thinner convective envelopes, observable signatures result from 10 $M_{\oplus}$ engulfment, and last for $\sim$1 Myr$-$2 Gyr.
Thus, engulfment signature detection in stars older than 2 Gyr assuming near-instantaneous nominal 10 $M_{\oplus}$ engulfment near ZAMS will be rare. 

The lifetimes of observable engulfment signatures can increase under different accretion scenarios, or with different engulfing star parameters. Late-stage engulfment of a 10 $M_{\oplus}$ planet by a 1 $M_{\odot}$ star at 300 Myr$-$3 Gyr post-ZAMS produces observable enrichment on timescales of $\sim$1.5 Gyr. For comparison, engulfment under the same conditions but occurring exactly at ZAMS results in signatures that remain observable for only $\sim$90 Myr. Lowering the engulfing star metallicity also increases engulfment signature lifetimes; 1 $M_{\odot}$ models with sub-solar metallicities ($Z$ = 0.012) sustain $>$0.05 dex levels of enrichment for $\sim$3 Gyr following engulfment of a 10 $M_{\oplus}$ planet. Thus, engulfment signatures in stars older than 1.5 Gyr are more likely to be observed if the star is low metallicity, and/or if engulfment occurred late in the lifetime of the system.

\subsection{Comparison to Observations}
As mentioned in Section \ref{sec:intro}, there are several reported binaries with $>$0.05 dex refractory differences. All the potentially engulfing stars in these systems are in the solar-like mass regime, and those with reported ages are older than 1.5$-$2 Gyr. Late ages are typical for these systems; the strongest engulfment detection reported thus far is HD 240429-30 (Kronos-Krios), which is 4.0$-$4.28 Gyr old \citep{oh2018}. Such refractory enrichments are unlikely to stem from ZAMS engulfment because our \texttt{MESA} models do not exhibit observable engulfment signatures on timescales longer than $\sim$2 Gyr for solar-like stars under these conditions. Late-stage (300 Myr$-$3 Gyr post-ZAMS) engulfment is also unlikely as signatures will persist on timescales of only $\sim$1.5 Gyr for 1 $M_{\odot}$ stars. However, four of the reported systems have sub-solar metallicities ($\zeta^{2}$ Reticuli, HD 134439$-$40, HIP 34407$-$26, and HD 133131) below $Z$ = 0.012, which increase signature timescales from nominal 10 $M_{\oplus}$ engulfment to $\sim$3 Gyr. 


The non-engulfing (Krios) and engulfing (Kronos) companions of Kronos-Krios have metallicities of [Fe/H] = 0.01 and 0.20 dex ($Z$ = 0.017 and 0.027), respectively. Thus, potential explanations for the strong signature (15 $M_{\oplus}$ of estimated engulfed material, \citealt{oh2018}) in this system could be some combination of late-stage engulfment, a large ($\gtrsim$50 $M_{\oplus}$) amount of mass engulfed, and chance observation of Kronos-Krios soon after its engulfment event. An alternative explanation is that the Kronos-Krios abundance differences are primordial rather than due to planet engulfment. This possibility is supported by the large projected separation of the system (11,000 $\pm$ 12 AU), which we calculated from Gaia DR3 astrometry. These lengths exceed typical turbulence scales in molecular clouds (0.05$-$0.2 pc, \citealt{brunt2009} and references therein), indicating that Kronos-Krios may have formed in distinct areas of chemodynamical space within their birth cloud.

Stellar twin pairs ($\Delta T_{\textrm{eff}} < 200$ K, \citealt{andrews2019}) are best for uncovering engulfment signatures because their abundances are free from differences due to mass-dependent evolution. However, many stellar binaries are composed of stars with different masses and stellar types. We found that higher mass stars ($M \gtrsim 1 \, M_\odot$) with thinner convective zones exhibit faster refractory depletion following engulfment (Figure \ref{fig:figure1}). \citet{sevilla2022} found this can be true even in the absence of engulfment due to gravitational settling. Thus, binary pair stars with different masses that begin chemically homogeneous may have different surface abundances at late times, even without planet engulfment. We thus argue that engulfment signatures are most reliable for twin binaries with nearly equal mass components, and observations targeting planet engulfment should focus on such systems.


Our models indicate that engulfment signatures are longest lived in stars with $M$ $\approx$ 1.1$-$1.2 $M_{\odot}$ (ZAMS temperatures of $T_{\rm eff}$ $\approx$ 6100$-$6400 K) and are thus more likely to be observed in stars slightly hotter than the Sun. 
We also predict longer-lived signatures in lower metallicity stars, which are hotter at the same mass. While the primordial metallicity (i.e., the metal content not including the engulfed planetary material) of stars is not easy to determine, we predict that stars with observable engulfment signatures will have lower primordial metallicity on average than non-enriched stars. This prediction could be tested with abundance measurements of non-refractory material (e.g., carbon or nitrogen) to estimate the pre-engulfment metallicity of the star.

\subsection{Limitations} \label{sec:limitations}
There are a few limitations in our \texttt{MESA} implementation that could be addressed in future studies. One issue is how we implemented mixing beyond thermohaline, overshoot, diffusion, and radiative levitation to account for other poorly understood mixing processes that operate within stellar interiors, e.g., turbulence and rotationally induced mixing. We used a constant \texttt{min\_D\_mix} coefficient to add this extra mixing in order to match observations of lithium depletion in solar-like stars \citep{sevilla2022}. However, this simple prescription will not reproduce the higher amount of mixing expected below the convective zones of massive stars due to, e.g., meridional circulation. Radiative levitation will also play a more prominent role in hotter and more massive stars. More accurate prescriptions for additional mixing processes would be a good addition to future studies of planet engulfment.

As discussed in Section \ref{sec:bulk_earth_accretion}, we were unable to implement instantaneous accretion representing planet engulfment because \texttt{MESA} failed to converge. Instead, we implemented engulfment via rapid accretion, resulting in an engulfment duration of slightly shorter than 10,000 years. While this period is much shorter than timescales of refractory depletion ($>$1 Myr), it is still much longer than instantaneous engulfment, which should occur within years. In addition, we did not consider the scenario of rapid engulfment that could plunge an engulfed planet deep within the stellar interior, and lead to its slow disintegration within the star \citep{jia2018}. This could even result in the planet plunging past the convective zone in more massive stars, significantly decreasing the strength and duration of refractory enhancements in the photosphere. These issues should be taken into consideration to more accurately model different engulfment scenarios.

\section{Summary} \label{sec:summary}
We used \texttt{MESA} models to simulate planet engulfment and mixing of material throughout the star due to convection, thermohaline mixing, diffusion, gravitational settling, and radiative levitation. We examined the evolution of measurable surface abundances under a range of accretion scenarios by varying the engulfed planet mass, the properties of the engulfing star, and the engulfment time and/or duration (pre-MS disk accretion, ZAMS, late heavy bombardment, and late-stage). We found that these conditions greatly affect the strength and duration of resultant planet engulfment signatures. 

Near-instantaneous engulfment occurring when the engulfing star reaches ZAMS results in different timescales for observable engulfment signatures as a function of engulfing star mass. At solar metallicity, the signature is largest and longest-lived for stars with $M$ $\approx$ 1.1$-$1.2 $M_{\odot}$.
We found that following planetary engulfment, thermohaline mixing quickly depletes the engulfment signature, lowering the increased surface abundances by a factor of $\sim$2 in the first $\sim$5$-$45 Myr. Afterwards, gravitational settling further depletes surface abundance enhancements on longer timescales. Observable signatures last for less than $\sim$2 Gyr in all our models, apart from our $1.2 \, M_\odot$ model where radiative levitation causes a $\sim$5 Gyr post-engulfment increase in surface abundances of certain chemical species (e.g., $^{56}$Fe, $^{27}$Al, $^{28}$Si). Engulfment scenarios mimicking LHB or pre-MS disk accretion result in shorter observable signature timescales compared to ZAMS engulfment. In LHB engulfment, refractory material is continuously depleted throughout longer accretion periods, resulting in almost no refractory enrichment post-engulfment. In disk accretion, refractory material is ingested by the star earlier in its lifetime, and is more heavily diluted within its larger convective envelope (relative to the size of a convective envelope of a star at ZAMS). This results in signatures weaker than those of ZAMS scenarios, but slightly stronger than those of LHB scenarios.  

Observable engulfment signature lifetimes increase for late-stage engulfment (300 Myr$-$3 Gyr post-ZAMS) scenarios, where the signatures of $10 \, M_\odot$ engulfment can persist for $\sim$1.5 Gyr in solar-like stars. Observable signatures are also more prominent for low-metallicity stars, lasting for $\sim$3 Gyr in $1\, M_\odot$ stars with $Z$ = 0.012 and ZAMS accretion. These conditions are thus more likely explanations for engulfment signatures observed in solar-like stars with ages $>$1.5 Gyr.
    
The strong dependence of engulfment signature strength and duration on stellar type, along with fewer theoretical uncertainties when modeling equal-mass stars, both underscore that stellar twin binaries are best-suited for observational planet engulfment surveys. We conclude that twin binaries with 1.1$-$1.2 $M_{\odot}$ masses are the most promising targets for engulfment detections.

\section*{Acknowledgments}
A.B. acknowledges funding from the National Science Foundation Graduate Research Fellowship under Grant No. DGE1745301. This work benefited from involvement in ExoExplorers, which is sponsored by the Exoplanets Program Analysis Group (ExoPAG) and NASA’s Exoplanet Exploration Program Office (ExEP).
\software{\texttt{numpy} \citep{numpy}, \texttt{matplotlib} \citep{matplotlib}, \texttt{pandas} \citep{pandas}, \texttt{scipy} \citep{scipy}, \texttt{astropy} \citep{astropy:2013, astropy:2018}}

\section*{Data Availability}
The data underlying this article will be uploaded to Zenodo.org  at \url{https://zenodo.org/communities/mesa} upon acceptance for publication.

\bibliography{mybib}{}
\bibliographystyle{aasjournal}



\end{document}